# Mid IR single-cycle light bullet self-reconstruction after an air gap under single-pulse femtosecond filamentation in LiF

**Sergey Chekalin,**[1] **Alexander Dormidonov,**[1,*] **Valerii Kandidov,**[2,1] **Victor Kompanets**[1]

[1] *Institute of Spectroscopy RAS, Moscow, Troitsk, 108840 Russia*
[2] *Moscow Lomonosov State University, Physics Department, Moscow, 119991 Russia*
*Corresponding author: dormidonov@gmail.com*

**The results of investigation of extremely compressed wave packet — 'light bullet' (LB) penetration through an air gap under single-pulse femtosecond Mid IR filamentation in LiF are presented. It is revealed by the laser coloration method and from numerical simulations that a single-cycle LB, which was formed before an air gap up to 0.5 mm width, completely recovered after passing some distance in LiF after the gap. This distance increases nonlinearly with the gap width and LB pathway before the gap. LB in the air gap has a strongly convergent wave front with focusing radius of 20 – 100 µm and its divergence after the waist is considerably less than that of Gaussian beam. © 2019 Optical Society of America**

http://dx.doi.org/10.1364/OL.99.099999

Self-reconstruction is a remarkable feature of light filaments allowing them to be transmitted through strongly turbulent and scattering media without an apparent change of the spatial profile and may be very important for various atmospheric, technological and biomedical applications [1-12]. This phenomenon, extensively investigated for femtosecond laser pulses filamentation in the near IR spectral range corresponding to the normal group velocity dispersion (GVD) has been ascribed to reshaping of the Gaussian input beam into a 'Bessel-like' beam [6,7] and explained by the existence of an extended low intensity periphery energy reservoir surrounding a narrow high-intensity central core of filament, which structurally sustains and continuously refills the central spike along the filament propagation [13]. In many works it was shown that filaments were terminated after blocking the periphery part of the filament except its core, but reconstructed after blocking the core itself [9-12]. In addition, in the normal GVD regime a strong energy interchange between the near-axis and peripheral parts of the filament [14] results in multiple refocusing [15], which is a repeating process of radiation self-focusing up to plasma formation, which leads to the intensity clamping and to the spatial divergence of the pulse tail.

At the regime of anomalous GVD a quite new filamentation regime occurs when extremely compressed wave packets up to one-optical cycle — so called light bullets (LBs) [16] have been observed [17-21]. There is only one experimental work on LB reconstruction [22], where LB penetration through an air gap in sapphire has been investigated. The observed capability of LB to restore its parameters after the gap provided the support of conception of the crucial role of an energy reservoir in this process once again. On the other hand in our experiments on filamentation in fused silica at anomalous GVD [17,20], when more than 90% of the periphery energy reservoir was blocked by 50-µm pinhole placed at the output face of a silica sample [23], from one to three successive few-optical cycle LBs have been recorded after passing several tens cm in air. It is in contrast with filament behavior in the case of normal GVD when the filamentary propagation is terminated after filament passing through a pinhole, which blocks the periphery and the transmitted central core quickly diffracts and decays.

In the present work we investigated LBs self-reconstruction experimentally and by numerical simulations. Specifically, LB propagation through an air gap and its subsequent recovering in LiF after the gap has been explored. The laser coloration technique [24 and references therein] based on the long-lived structures of color centers (CCs) formation by a single laser pulse is used to investigate this effect in LiF.

The experimental layout was similar to that described in [18]. The 130 fs pulse at 3200 nm with energy of about 25 µJ was focused by a thin CaF$_2$ lens with focal distance 97 mm inside LiF sample with an air gap that can be changed from 10 to 500 µm. At these experimental parameters, effective compression of LB up to a single optical cycle takes place at all stages of its propagation in LiF before as well as after the air gap. At that the LB formation threshold, which is determined by the ratio of dispersion and diffraction length [25], remains practically constant and is not more than stationary critical power for self-focusing.

The lengths of the parts of the LiF sample before and after the air gap were about 7 mm. To obtain a single-pulse exposure, the sample was translated after each shot in the direction perpendicular to the laser beam. To analyze the spatial distribution of the luminescence intensity of the recorded CCs we used the method of optical microscopy with illumination of the induced CCs by cw laser radiation at 450 nm. In contrast to [19,21,22] laser coloration technique made it possible, first, to carry out the measurements without signal accumulation, i.e. using only one laser pulse, and second, without any inaccuracies arising from the overlap of the radiations of supercontinuum, conical emission, and plasma channel. The long-lived structures made up of CCs, may be easily detected and investigated under subsequent continuous laser illumination in their absorption band near 450 nm. This technique allows observation of the formation of single-cycle LBs and to reveal that it does not change its parameters after propagation through an induced waveguide [17,24].

To analyze the light field transformation during LB formation, we obtained a numerical solution of the unidirectional equation for Mid IR femtosecond pulse propagation in LiF [26,27]. As an initial condition we took a spectrally limited wave packet with a Gaussian distribution of the electric field amplitude in time and space corresponding to those used in the experiment. The numerical simulation output included the evolution of the light field with a single-cycle LB propagation in LiF and the distribution of free electrons density in the laser-induced plasma, which was compared with the CCs traces recorded in the experiment.

Fig. 1 shows the trace of the LB in a continuous LiF sample without an air gap. The experimentally registered LB starts after passing about 7 mm in LiF and the length of CCs trace is equal to (0.7±0.1) mm. The CCs density oscillations are caused by the periodic change of the light field intensity (red curve in Fig. 1) due to a difference between the phase velocity of the light field and the group velocity of the wave packet envelope during LB propagation — so called LB 'breathing' [17,18]. Numerically calculated electron density distribution in a plasma channel $N_e(x)$ was integrated by one transverse direction for accordance with experimental scheme of CCs observation and is fully reproducing the experimental pattern.

The blue curve in Fig. 1 shows the on-axis electron density profile $N_e(x=0)$ and the red curve — maximal on-axis intensity of the LB's light field. The CCs formation and, respectively, laser plasma threshold can be estimated from Fig. 1 and appears to be about $1.7 \cdot 10^{14}$ W/cm$^2$, while the maximal intensity in the LB does not exceed $2.0 \cdot 10^{14}$ W/cm$^2$. So the LB leaves a trace in the LiF sample when intensity is clamped in very narrow range. Vertical lines '1', '2', '3' in Fig. 1 marked the relative positions of an air gap within LB pathway in the first part of the sample, and were used for numerical simulation of LB reconstruction after the gap presented below.

Some examples of experimental data are presented in Fig. 2 for air gap values of 100 and 500 μm. It can be seen that CCs traces produced by different laser pulses are similar: the length of each trace (corresponding to LB path length) does not exceed (0.7±0.1) mm and the luminescence intensity along a trace is modulated that indicates the formation of a single-cycle LB [17,18,28] before as well as after the air gap.

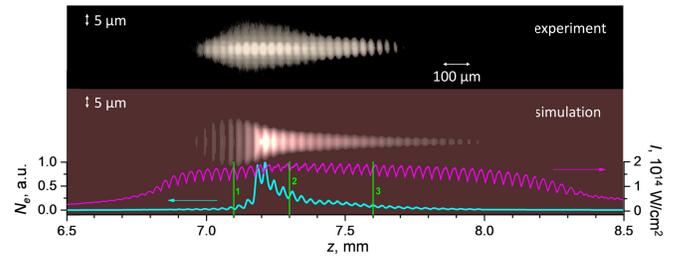

Fig. 1. The trace of the LB in the continuous LiF sample without an air gap: experiment (top) and numerical simulation (bottom). Blue curve — on-axis electron density, pink curve — on-axis LB peak intensity.

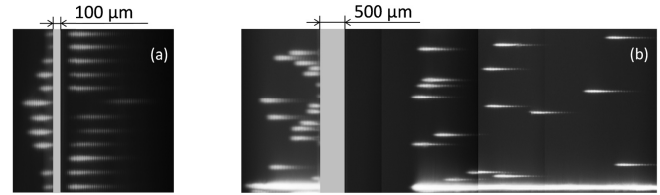

Fig. 2. Photos of the luminescence intensity distribution of the CCs structures induced by a single-cycle LB under filamentation of a single 130-fs pulse at a wavelength of 3200 nm in LiF with an air gap 100 μm (a) and 500 μm (b). The lowest trace in (b) was obtained under many pulses exposure. The pulse propagated from left to right. The vertical scale was stretched by a factor of 10 for the sake of clarity.

In spite of unchangeable laser pulse parameters the positions of CCs traces induced by a single pulse are rather different from shot to shot, and under many pulses exposure these traces are superimposed and form much more long blurred trace (Fig. 2b, bottom trace). This effect results in considerable inaccuracies in experiments with signal accumulation from many laser pulses (for instance in [19,21,22]). From Fig. 2 one can conclude that the reconstruction after an air gap was observed only for LBs, which did not passed through the full path length before the gap in the first part of the sample, otherwise LBs does not observed after the gap. In addition, it can be seen that the distance, which is needed for LB reconstruction after the air gap, is increases with the gap width.

Longitudinal on-axis luminescence intensity profiles for the CCs structures induced in LiF by single 130-fs pulses and recorded in experiments for air gaps of 25, 100, 390, and 500 μm and for different LB pathways before the gap are presented in the left side of Fig. 3. Corresponding calculated electron density distributions $N_e(x=0)$ (blue curves) and LB light field on-axis peak intensity (red curves) are shown in the right side of Fig. 3. The numbers '1', '2', and '3' indicate the relative positions of an air gap marked in Fig. 1.

The results of numerical simulations are consistent with experimentally measured CCs density profiles in LiF in spite of different nature of these signals, which manifested by slightly mismatch in their envelopes. Both the experimentally measured CCs density and the electron concentration in the laser induced plasma channel obtained from numerical simulations periodically oscillate with the distance. These oscillations indicate the formation and 'breathing' of the single-cycle LB before and after the air gap.

It should be pointed out that LB reconstruction after the air gap occurs at a distance that depends on the gap width and the LB

pathway before the gap. It can be seen that the 25-μm gap (Fig. 3a) practically has no influence on the formation and propagation of LB except of some reflection losses at the sample faces in the gap. With the gap width increasing the picture begins to change. First, just after the gap a region without plasma electrons and CC luminescence arises and only after this 'dark space' the LB reconstruction occurs. Minimal length of the distance needed for LB reconstruction which achieved for its minimal path length before the gap (position '1' in Fig. 1) increases from 50 μm for 100-μm gap (Fig. 3b) up to 1000 μm and 1300 μm for 390-μm (Fig. 3c) and 500-μm gap (Fig. 3d), respectively. At that several times increasing of the LB reconstruction distance was observed when larger part of LB pathway is located in the LiF sample before the air gap. In this case reconstruction distance increases to 2300 μm for 390-μm gap (Fig. 3c) and 3600 μm for 500-μm (Fig. 3d).

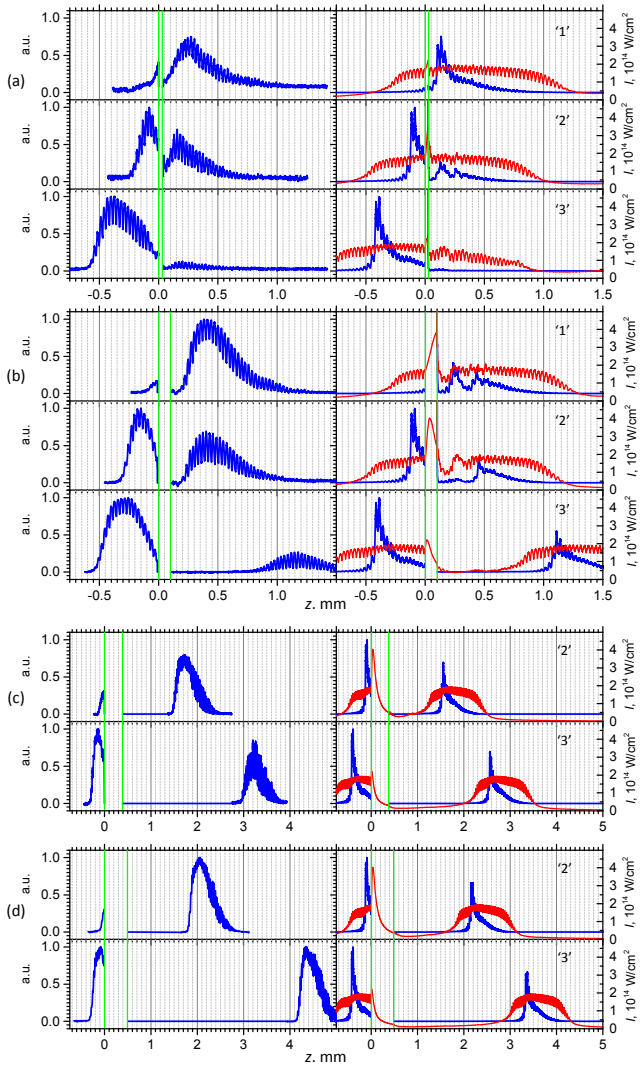

Fig. 3. Longitudinal luminescence intensity profiles for the CCs structures in LiF recorded in experiments (left). Calculated electron density distribution in a plasma channel (blue curves) and on-axis peak intensity (red curves) during the propagation of a single-cycle LB at the same conditions (right). The air gap width 25 μm (a), 100 μm (b), 390 μm (c), and 500 μm (d).

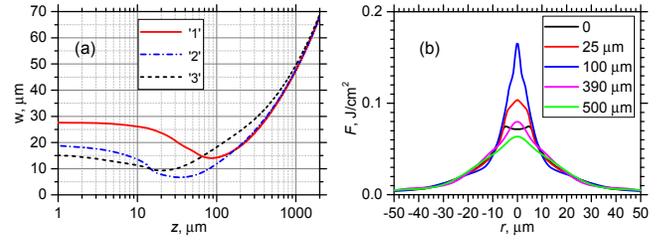

Fig. 4. (a) — LBs fluence transverse radius (at the $1/e^2$ level) in the air ('1', '2', '3' — positions of the gap within LB pathway in the sample); (b) — Fluence profiles at several distances z in air for the LB pathway '1'.

The origin of the reconstruction distance can be explained by the LB diffractive divergence in the process of its propagation in the air gap where there is no enough Kerr nonlinearity. The divergence leads to intensity decreasing after the gap on the input face of the second part of the sample up to a value less than plasma and CC formation threshold.

In Fig. 4a the calculated evolutions of LBs transverse radius (at the $1/e^2$ fluence level) with distance $z$ in the air for different positions ('1', '2', and '3') of the gap within LB pathway are presented. Numerical simulations show that LB just after the first part of the LiF sample in the air gap has a strongly convergent wave front and its curvature depends on the LB pathway before the gap. Therefore, this pathway determines the location of the beam waist inside the air gap and the beam size and LB intensity at the input face of the second part of the LiF sample after the air gap. For instance, after 10 and 25-μm air gaps LB diameter does not increase and practically does not affect the reconstruction of the LB, regardless of its pathway before the gap.

For the 100-μm gap when a short part of LB pathway is located in the LiF sample before the air gap (Fig. 3b, '1') the LB peak intensity in the beam waist reaches value more than $4 \cdot 10^{14}$ W/cm² near the input face of the second part of the LiF sample but quickly decays due to strong losses in a thin dense plasma layer arising here. In the cases of 390-μm and 500-μm gaps when LB diverges significantly in air the reconstruction of the LB after the gap occurs as a result of a nonlinear optical field compression. With an increase in the gap length, due to optical field divergence LB diameter at the input face of the second part of the sample increases, the intensity drops and the length of the compression region before LB reconstruction increases like the distance of the stationary self-focusing of diverging beams. By means of Fig. 4a data linear fitting we received the LB full divergence in far field zone is equal to 0.037 rad that about four times less than the divergence of Gaussian beam with the same waist. Our estimations are in good agreement with experimentally registered sub-diffractive LB propagation in the air after sapphire sample [21].

LB's transverse fluence distribution profiles for several distances $z$ in the air are presented in Fig. 4b for the case '1' (see Fig.1). It should be noted that the shapes of all these profiles differ from Bessel- or Gaussian-like. It can be seen that just after the exit from LiF LB has a ring-shaped fluence, which at the distance of 25 μm collapses to the on-axis maximum.

The on-axis LB peak intensity oscillations (red curves in Fig. 3) indicate the formation and 'breathing' of the single-cycle LB before and after the air gap. But due to equality of the phase and group velocities in the air the LB 'breathing' becomes freezing in the gap and a single-cycle LB's electric field temporal profile is nearly unvarying except an amplitude change caused by diffractive optical field divergence (Fig. 5). The high-frequency oscillations of electric field in the LB rear part represent a short-wavelength anti-Stocks wing of supercontinuum, which exhaustive practically all LB energy during its full pathway [17,23].

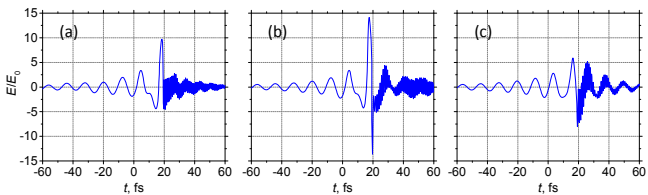

Fig. 5. LB's on-axis electric field temporal profiles in air at several distances $z$: 0 μm (a), 100 μm (b), and 500 μm (c).

Our investigations showed that single-cycle LBs are extremely robust formations and are capable to the complete spatiotemporal self-reconstruction in nonlinear dispersive media after propagation through an air gap up to 500 μm in linear (free-space) regime without dispersion that is in accordance with results of [22]. Application of laser coloration method in our experiments made it possible, in contrast to [19,21,22] to carry out the measurements without signal accumulation, i.e. using only one laser pulse. It allowed us, first, to reveal once more that LB pathway in nonlinear dispersive media is about (0.7±0.1) mm in accordance with our earlier works [17,18,24,25,28]. Second, we observed that only LB, which has not completely passed through their full pathway before the air gap, is capable to self-reconstruction after the gap. Third, it was revealed that reconstruction occurs at some distance in nonlinear media that depends on the gap width and the LB pathway before the gap. The origin of this reconstruction distance was derived from numerical simulations, which showed that LB in the air gap has a strongly convergent wave front and its curvature depends on LB pathway before the gap. Therefore, the LB pathway before the gap determines the location of the beam waist inside the air gap and the beam size and LB intensity in nonlinear medium after the air gap. With an increase in the gap length, due to optical field divergence LB's diameter increases, the intensity drops and the length of the compression region before LB reconstruction increases. For small air gap with widths that are comparable to LB focusing radius, the peak intensity in the beam waist can achieves more than $4 \cdot 10^{14}$ W/cm$^2$ near the input face of nonlinear medium after the gap. This leads to additional strong losses in a thin dense plasma layer arising here that also extends the length of the compression region. Our numerical simulations showed that the shape of LB's transverse fluence distribution profiles differs from Bessel- or Gaussian-like shape. At that data linear fitting revealed that LB full divergence in far field zone is about four times less than the divergence of Gaussian beam with the same waist. This result is in a good agreement with experimentally registered sub-diffractive LB propagation in the air after sapphire sample [21]. One more numerical result is that single-cycle LB's electric field temporal profile is nearly unvarying under propagation without dispersion in air gap. This feature of LB can be used for ultrafast diagnostic applications when a single-cycle probe pulse is needed.

**Funding.** Russian Science Foundation (18-12-00422).

**Disclosures**. The authors declare no conflicts of interest.